\documentclass{iau}

\usepackage{amsmath}
\usepackage{graphicx}
\usepackage{multirow}
\usepackage{pgf,tikz}
\usepackage{caption}
\definecolor{bleufonce}{rgb}{0.35, 0.0, 0.55} 
 
 \def\fds{\hbox{$.\!\!{''}$}}
\definecolor{carotte}{rgb}{0.96, 0.4,0.11}

\begin{document}

\lefttitle{Rosu et al.}
\righttitle{$\tau$ CMa: the hierarchical quadruple system}

\jnlPage{1}{7}
\jnlDoiYr{2025}
\doival{}

\aopheadtitle{Proceedings IAU Symposium}
\editors{A. Wofford,  N. St-Louis, M. Garcia \&  S. Simón-Díaz, eds.}

\title{One century data of $\boldsymbol{\tau}$ CMa: a (2+1)+1 system with a short-period overcontact binary and an eccentric intermediate orbit with probably no apsidal motion}

\author{S.~Rosu$^1$, J.~Maíz Apellániz$^2$, L.~Sciarini$^1$, R.~C.~Gamen$^{3,4}$, J.~A.~Molina-Calzada$^{2,5}$, G.~Holgado$^6$, \& R.~H.~Barbá$^7$}
\affiliation{$^1$Département d’Astronomie, Université de Genève, Chemin Pegasi 51, CH-1290 Versoix, Switzerland.\\
$^2$Centro de Astrobiología, CSIC-INTA, Campus ESAC, C. bajo del castillo s/n, E-28 692 Villanueva de la Cañada, Madrid, Spain.\\
$^3$Instituto de Astrofísica de La Plata, CONICET–UNLP, Paseo del Bosque s/n, La Plata, Argentina. $^4$Facultad de Ciencias Astronómicas y Geofísicas, UNLP, Paseo del Bosque s/n, La Plata, Argentina.\\
$^5$Departamento de Astrofísica y Física de la Atmósfera, Universidad Complutense de Madrid, E-28 040 Madrid, Spain.\\
$^6$Instituto de Astrofísica de Canarias, Universidad de La Laguna, Tenerife, Spain.\\
$^7$Departamento de Astronomía, Universidad de La Serena, Av. Juan Cisternas 1200 Norte, La Serena, Chile.}

\begin{abstract}
$\boldsymbol{\tau}$ Canis Majoris (CMa) is an intriguing system that has captured astronomers’ attention for more than a century.
The two main components Aa and Ab are two evolved O stars on a 350 years orbit. Aa is itself a SB1 with a 155-days period and a 0.3 eccentricity. 
Since Hipparcos, we know that a 1.28-days period eclipsing binary (EB) is hidden somewhere in Aa or Ab, but nowhere else.
Our recent analysis finally disentangles the system.
We calculated the visual Aa--Ab orbit from AstraLux imaging.
We detected the SB2 nature of Aa based on STIS spectra, the companion of the O star (Aa1) being a B+B binary (Aa2~$=$~Aa2a~$+$~Aa2b). Multiple lines of evidence point towards Aa2 being the EB: time delays in the eclipsing orbit detected by TESS, high mass for Aa2 from SB1 from constraints from the orbit of Aa1, and a lack of radial-velocity motion of Ab synchronised with the eclipsing orbit. This remains as a tentative conclusion pending further analysis.
We detect secular changes in the SB1 orbit of Aa1 on a baseline longer than a century. At this stage, the effect is most likely caused by the change in velocity of the Aa center of mass due to the Aa--Ab visual orbit. Apsidal motion is most probably not the culprit.

\end{abstract}


\maketitle

\section{$\boldsymbol{\tau}$ CMa: Stellar multiplicity does not have to be simple}
\noindent$\tau$ CMa, located at the center of NGC\,2362, is among the 15 brightest O-type systems in our sky. Its multiplicity is complex, as illustrated in Fig.\,\ref{fig:multiplicity}. We combined astrometry (Sect.\,\ref{sect:astrometry}), spectroscopy (Sect.\,\ref{sect:spectroscopy}), photometry (Sect.\,\ref{sect:photometry}), and three-body simulations (Sect.\,\ref{sect:tres}) to disentangle $\tau$ CMa A's, though more is to come (Sect.\,\ref{sect:conclusion}).  

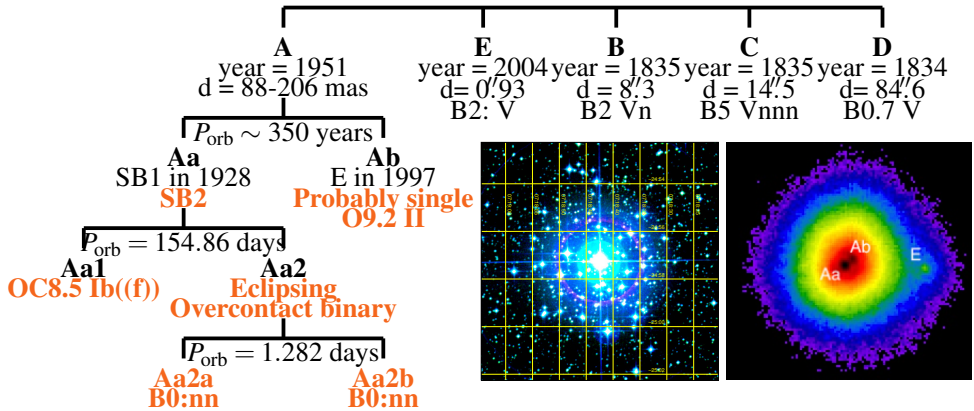
\begin{figure}
\label{fig:multiplicity}
\begin{tikzpicture}[scale=0.44]
\draw[line width=0.5mm] (0,-0.4)--(18,-0.4);
\draw[line width=0.5mm] (0,-0.4)--(0,-1);
\draw[line width=0.5mm] (6,-0.4)--(6,-1);
\draw[line width=0.5mm] (10,-0.4)--(10,-1);
\draw[line width=0.5mm] (14,-0.4)--(14,-1);
\draw[line width=0.5mm] (18,-0.4)--(18,-1);
\draw[line width=0.5mm] (0,-3.1)--(0,-3.7);
\draw[line width=0.5mm] (-3,-3.7)--(3,-3.7);
\draw[line width=0.5mm] (-3,-3.7)--(-3,-4.3);
\draw[line width=0.5mm] (3,-3.7)--(3,-4.3);
\draw[line width=0.5mm] (-3,-6.4)--(-3,-7.0);
\draw[line width=0.5mm] (-6,-7.0)--(0,-7.0);
\draw[line width=0.5mm] (-6,-7.0)--(-6,-7.6);
\draw[line width=0.5mm] (0,-7.0)--(0,-7.6);
\draw[line width=0.5mm] (0,-9.7)--(0,-10.3);
\draw[line width=0.5mm] (-3,-10.3)--(3,-10.3);
\draw[line width=0.5mm] (-3,-10.3)--(-3,-10.9);
\draw[line width=0.5mm] (3,-10.3)--(3,-10.9);

\draw[below](0,-1)node{\textbf{A}};
\draw[below](6,-1)node{\textbf{E}};
\draw[below](10,-1)node{\textbf{B}};
\draw[below](14,-1)node{\textbf{C}};
\draw[below](18,-1)node{\textbf{D}};

\draw[below](0,-1.6)node{year = 1951};
\draw[below](6,-1.6)node{year = 2004};
\draw[below](10,-1.6)node{year = 1835};
\draw[below](14,-1.6)node{year = 1835};
\draw[below](18,-1.6)node{year = 1834};

\draw[below](0,-2.2)node{d = 88-206 mas};
\draw[below](6,-2.2)node{d= 0\fds93};
\draw[below](10,-2.2)node{d = 8\fds3};
\draw[below](14,-2.2)node{d = 14\fds5};
\draw[below](18,-2.2)node{d= 84\fds6};

\draw[below](6,-2.9)node{B2: V};
\draw[below](10,-2.9)node{B2 Vn};
\draw[below](14,-2.9)node{B5 Vnnn};
\draw[below](18,-2.9)node{B0.7 V};

\draw[](0,-4.2)node{$P_\text{orb} \sim$ 350 years};
\draw[below](-3,-4.3)node{\textbf{Aa}};
\draw[below](-3,-4.9)node{SB1 in 1928};
\draw[below](-3,-5.5)node{\textcolor{carotte}{\textbf{SB2}}};

\draw[below](3,-4.3)node{\textbf{Ab}};
\draw[below](3,-4.9)node{E in 1997};
\draw[below](3,-5.5)node{\textcolor{carotte}{\textbf{Probably single}}};
\draw[below](3,-6.1)node{\textcolor{carotte}{\textbf{O9.2 II}}};

\draw[](-3,-7.5)node{$P_\text{orb} = 154.86$ days};

\draw[below](-6,-7.6)node{\textbf{Aa1}};
\draw[below](-6,-8.2)node{\textcolor{carotte}{\textbf{OC8.5 Ib((f))}}};

\draw[below](0,-7.6)node{\textbf{Aa2}};
\draw[below](0,-8.2)node{\textcolor{carotte}{\textbf{Eclipsing}}};
\draw[below](0,-8.8)node{\textcolor{carotte}{\textbf{Overcontact binary}}};

\draw[](0,-10.8)node{$P_\text{orb} = 1.282$ days};
\draw[below](-3,-10.9)node{\textcolor{carotte}{\textbf{Aa2a}}};
\draw[below](-3,-11.5)node{\textcolor{carotte}{\textbf{B0:nn}}};
\draw[below](3,-10.9)node{\textcolor{carotte}{\textbf{Aa2b}}};
\draw[below](3,-11.5)node{\textcolor{carotte}{\textbf{B0:nn}}};

\node (im1) at (9.5,-8) {\includegraphics[width=90pt]{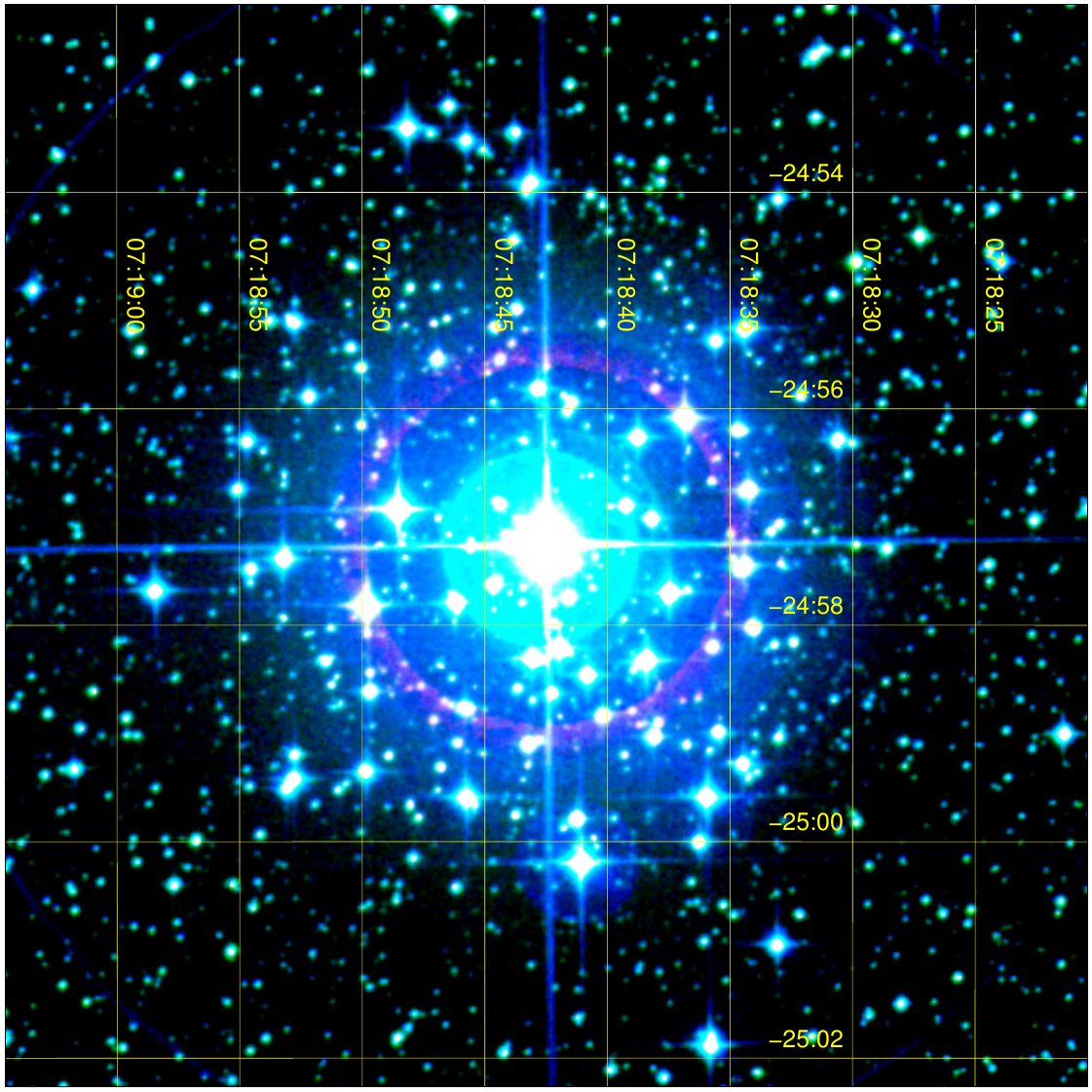}};
\node (im2) at (17,-8.01) {\includegraphics[clip=true, trim=240 360 240 370,width=92pt]{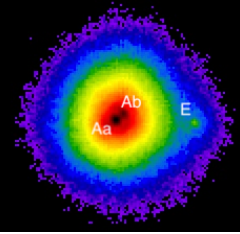}};


\end{tikzpicture}
\caption{Hierarchy of $\tau$ CMa. Results of our analysis are in orange. First image: 3 channel DDS2 RGB image. Second image: AstraLux $3"\times3"$ cut-out of the region around $\tau$ CMa A.}
\label{fig:multiplicity}
\end{figure}

\section{Astrometry: Long visual outer orbit Aa--Ab\label{sect:astrometry}}
\noindent We combined data from \citet{maiz20} with new AstraLux and STIS data.

\begin{figure}
\begin{minipage}[b]{0.56\linewidth}
\centering
\includegraphics[clip=true,trim=0 0 0 0,width=0.9\linewidth]{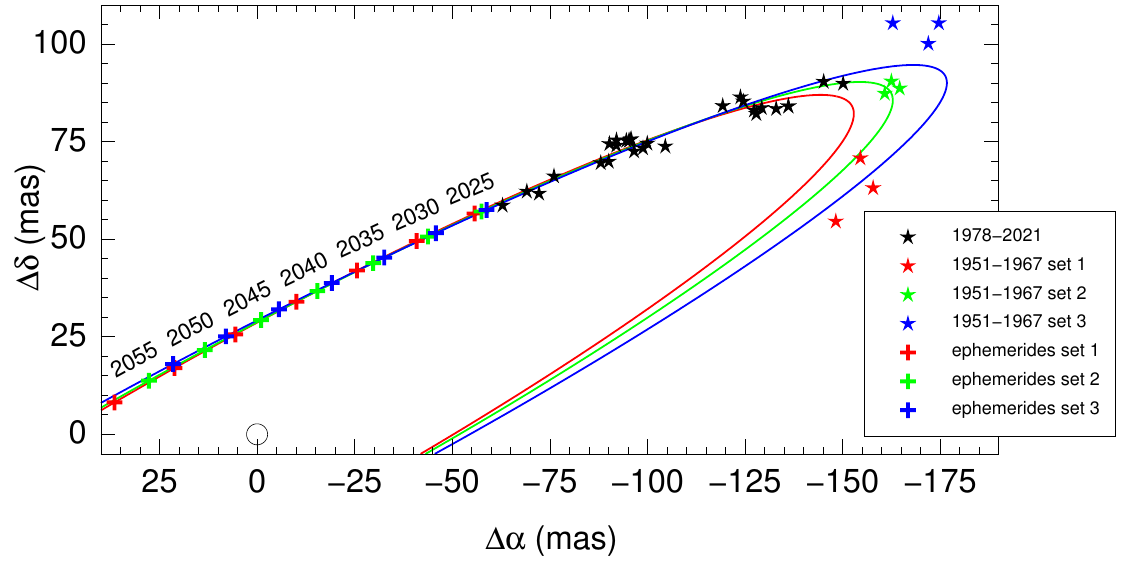}
\caption{Visual orbit of Aa--Ab. Colours indicate three possible orbits based on different sets of data.}
\label{fig:astra}
\end{minipage}
\hfill
\begin{minipage}[b]{0.34\linewidth}
\centering
\includegraphics[clip=true,trim=0 0 0 0,width=0.40\linewidth]{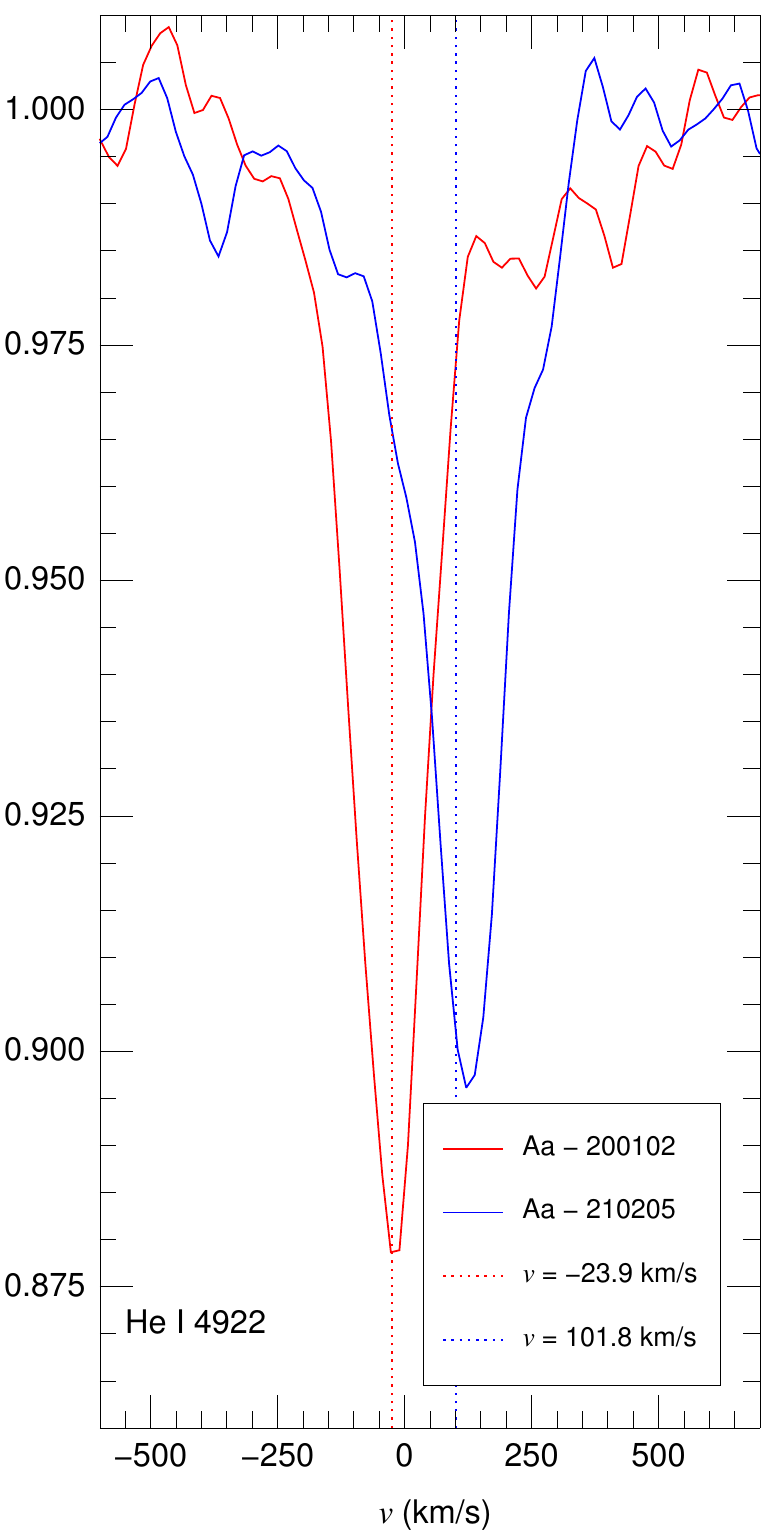}
\includegraphics[clip=true,trim=0 0 0 0,width=0.40\linewidth]{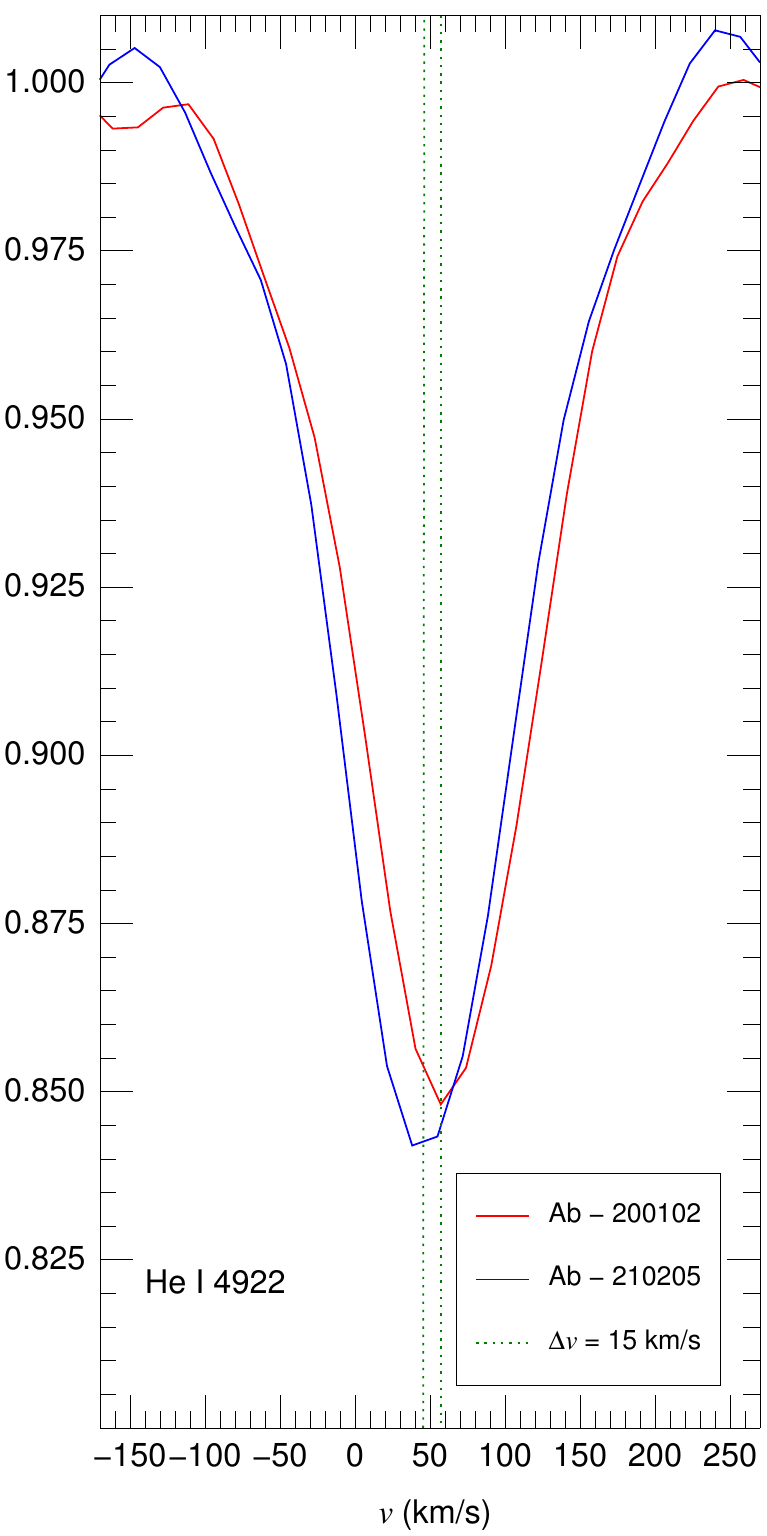}
\caption{STIS spectroscopy: Aa (left) and Ab (right).}
\label{fig:STIS}
\end{minipage}
\end{figure}

\noindent\textbf{AstraLux lucky imaging.} Data taken between 1951-1967 are incompatible with each others. We thus computed three possible orbits assuming $e=0$ (see Table\,\ref{table:astra} and Fig.\,\ref{fig:astra}), and showed that the parameters $i, \Omega, M_\text{Aa,Ab}, K_\text{Aa,Ab}$ are well-constrained.

\noindent\textbf{HST/STIS spectroscopy.} Aa shows a broad component that moves in anti-phase with Aa1 (see Fig.\,\ref{fig:STIS}). Thus, Aa is a SB2, with Aa2 a fast rotator. We determine the mass of Ab $\sim\,25\pm5\,M_\odot$ from the cluster age and its O9.2\,\textsc{II} spectral type.

\vspace*{-0.8cm}

\setlength{\tabcolsep}{2pt}
\noindent \begin{minipage}[t]{0.5\linewidth}
\centering
\captionof{table}{Three possible orbital solutions for Aa--Ab.}
\begin{tabular}{c c c c c c c}
\hline
$P_\text{orb}$& $T_0$& $a$  & $i$ & $\Omega$ & $M_\text{Aa,Ab}$ & $K_\text{Aa,Ab}$    \\
(yr) & (yr) &  (mas) &  ($^\circ$) & ($^\circ$) & ($M_\odot$) &  (km\,s$^{-1}$)   \\
\hline
306.8 & 1970.4 & 174.1 & 81.7 & 298.9 & 103.5 & 20.5 \\
354.5 & 1960.9 & 184.5 & 82.1 & 298.3 &  92.3 & 18.8\\
408.7 & 1949.3 & 198.8 & 82.5 & 297.5 & 86.9 & 17.6\\
\hline
\end{tabular}
\label{table:astra}
\end{minipage}
\hfill
\begin{minipage}[t]{0.42\linewidth}
\centering
\captionof{table}{Orbital solution for Aa1--Aa2.\label{table:RV}}
\begin{tabular}{l l l}
\hline
Parameter & Value \\
\hline
$T_0$ (HJD) & $2\,455\,098.3  \pm 0.4$  \\
$\omega_0$ ($^\circ$) & $84.3 \pm 1.1 $ \\
$e$ & $ 0.280 \pm 0.005$ \\
$P_\text{orb}$ (d)& $154.900\pm 0.004$ \\
$K_P$ (km\,s$^{-1}$) & $87.0 \pm 0.5$ \\
$a_P\sin\,i$ ($R_\odot$) &  $1.189 \pm 0.007$ \\
$M_P\sin^3 i$ ($M_\odot$) &  $9.34 \pm 0.16$  \\
$\chi^2_\text{red}$ & 0.89\\
\hline
\end{tabular}
\end{minipage}

\section{Spectroscopy: Intermediate orbit Aa1--Aa2\label{sect:spectroscopy}}
\noindent \textbf{Radial velocity analysis.} 
We combined 125 RVs spanning 117 years that we analysed as in \citet{rosu20,rosu22a,rosu22b} to derive the orbital solution (Table\,\ref{table:RV}). Newly accumulated data were combined with data from \citet{struve28, struve54, stickland98}, and LiLiMaRlin \citep{maiz19}. 
Combining $M_\text{Aa,Ab}$ with constraints on $\sin i_\text{Aa1,Aa2}$ and $M_\text{Aa2}/M_\text{Aa1} \sim 1.0-1.5$, we deduced that the orbit is nearly edge on, and that $M_\text{Aa1}\sim 30\,M_\odot$, and $M_\text{Aa2}\sim 38\,M_\odot$. The only sign of Aa2 in the deconvolved spectra is under broad profiles in some lines, suggesting that Aa2 has early-B spectrum/spectra. Furthermore, the velocity amplitude of Aa2 is much lower than that of Aa1, thus $M_\text{Aa2} > M_\text{Aa1}$. It only works if Aa2 is made up of two early-B fast rotators of $\sim 19\,M_\odot$ each!

\noindent\textbf{Secular changes in the SB2 orbit of Aa1.}
Over the period 2004-2025, we detected a relative average acceleration of the center of mass of Aa with respect to Ab of $0.22\pm0.06$ km\,s$^{-1}$\,yr$^{-1}$ that is consistent with the expectations of the visual Aa--Ab orbit if it is circular (Sect.\,\ref{sect:astrometry}), as well as a mass ratio between Ab and Aa of $1.00\pm0.53$ consistent with the expected one (within the large error bar). Most likely, and to the first author dismay, there is no apsidal motion.

\section{Photometry: Short-period orbit Aa2a--Aa2b\label{sect:photometry}}
\begin{figure}
\begin{minipage}[t]{0.36\linewidth}
\centering
\includegraphics[clip=true,trim=35 5 40 25,width=\linewidth]{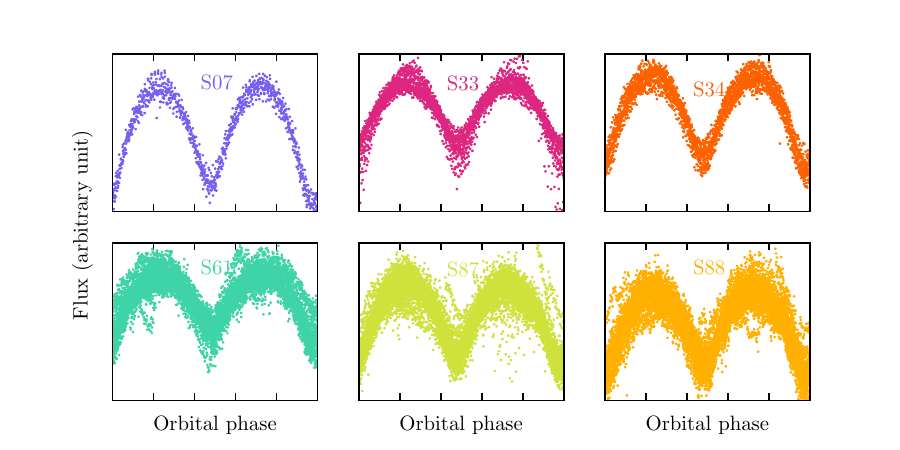}
\caption{TESS light curves of the eclipsing binary.}
\label{fig:tess}
\end{minipage}
\hfill
\begin{minipage}[t]{0.61\linewidth}
\centering
\includegraphics[clip=true,trim=10 0 40 40,width=0.42\linewidth]{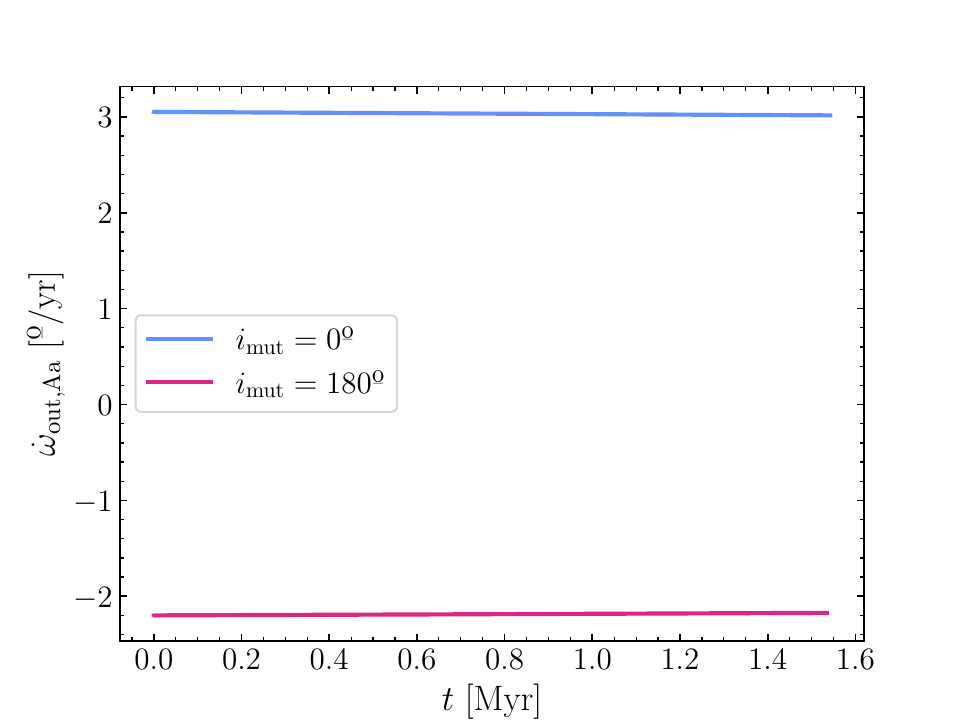}
\includegraphics[clip=true,trim=0 0 50 40,width=0.50\linewidth]{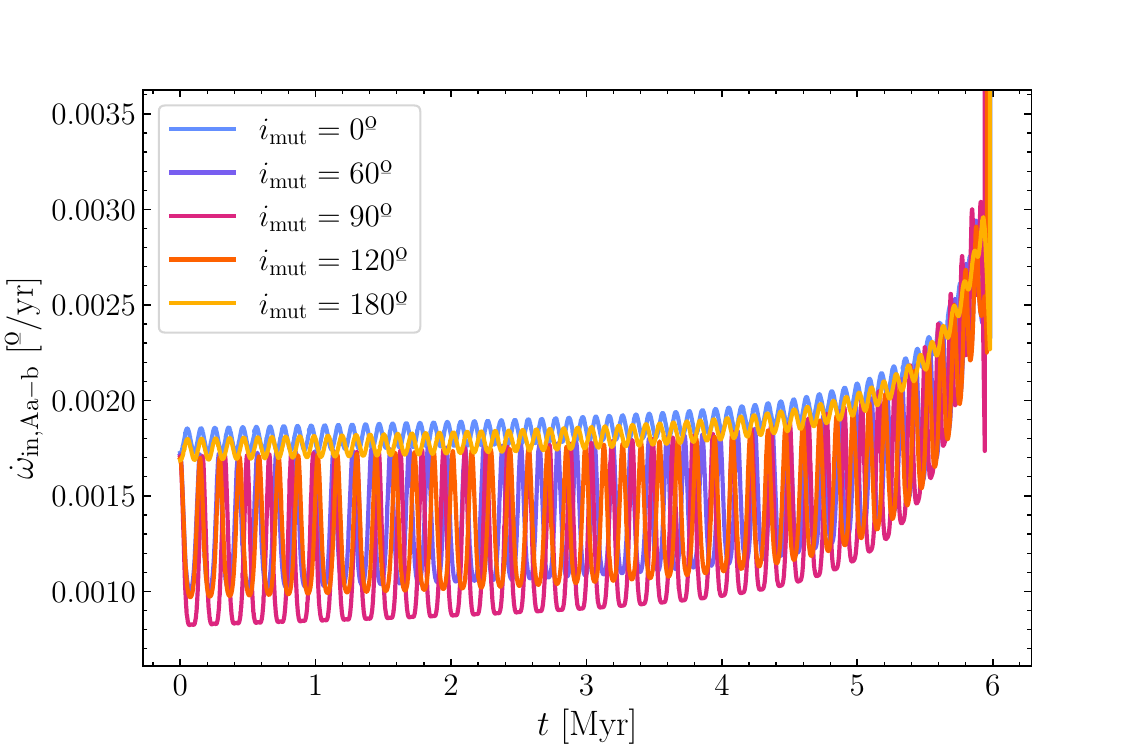}
\caption{TRES simulations: $\dot\omega$ in Aa1--Aa2 if induced by Aa2a--Aa2b (left) or Ab (right) for different $i_\text{mut}$.}
\label{fig:tres}
\end{minipage}
\end{figure}

\noindent TESS light curves of $\tau$ CMa reveal that the close eclipsing binary is an overcontact binary with a 1.282\,days period (Fig.\,\ref{fig:tess}). Detailed analysis is ongoing to derive stars and orbital parameters.

\section{Three-body simulations with TRES\label{sect:tres}}
\noindent We performed three-body simulations with TRES \citep{toonen16, sciarini25} to assess whether Aa2a--Aa2b (Case 1, Fig.\,\ref{fig:tres}, left panel) and/or Ab (Case 2, Fig.\,\ref{fig:tres}, right panel) could be (partly) responsible for an apsidal motion in Aa1--Aa2.\\
\textbf{Case 1.} We considered the triple system Aa1--(Aa2a--Aa2b). If the mutual inclination $i_\text{mut}$ between the two orbits is $0^\circ$, the apsidal motion rate $\dot\omega$ amounts to $\sim 3.1^\circ$\,yr$^{-1}$, while if $i_\text{mut} = 180^\circ$, $\dot\omega \sim -2.2^\circ$\,yr$^{-1}$. These are the two extreme cases; any other $i_\text{mut}$ leads to a value for $\dot\omega$ intermediate between 3.1 and $-2.2^\circ$\,yr$^{-1}$.\\
\textbf{Case 2.} We considered the triple system (Aa1--Aa2)--Ab. Whatever $i_\text{mut}$, $\dot\omega \le 0.002^\circ$\,yr$^{-1}$.\\
Should an apsidal motion be measured in Aa1--Aa2, would Aa2a--Aa2b be the culprit, not Ab.


\section{What we still do not know but will soon\label{sect:conclusion}}
\noindent Our analysis of $\tau$ CMa A allowed us to unravel its mystery. Yet, it remains to perform a quantitative spectroscopic analysis of its components Aa1, Aa2a, Aa2b, and Ab to derive the effective temperatures and luminosities of the stars as well as their surface chemical abundances and wind properties. Furthermore, the origin of the secular changes of the SB1 Aa1--Aa2 orbit as well as the properties of the overcontact binary Aa2 are still to be determined.

\end{document}